\documentclass{iaus}
\usepackage{graphicx}
\def\kms{\mbox{$\rm km~s^{-1}$}}
\def\sec{$^{\prime\prime}$~}

\title[short title of paper] %% give here short title %%
{Radio Observations of AGN in Low Surface Brightness Galaxies}

\author[]   %% give here short author list %%
{M.Das$^1$, S.S.McGaugh$^2$, N.Kantharia$^3$, S.N.Vogel$^2$}

\affiliation{$^1$Raman Research Institute, Bangalore 560 080, India;
email: mousumi@rri.res.in \break
$^2$University of Maryland, College Park, MD 20742\break
$^3$NCRA, TIFR, Post~Bag 3, Ganeshkhind, Pune-411007, India}

\pubyear{2007}
\volume{244}  %% insert here IAU Symposium No.
\pagerange{xxx--xxx}
\date{?? and in revised form ??}
\jname{Proceedings Title IAU Symposium}
\editors{xxx \& xxx, eds.}
\begin{document}

\maketitle

\vspace{-2mm}
\begin{abstract}
We present preliminary results of a study of the low frequency radio continuum 
emission from the nuclei of Giant Low Surface Brightness (LSB) galaxies. 
We have mapped the emission and searched 
for extended features such as radio lobes/jets associated with AGN activity. 
LSB galaxies are poor in star formation and generally less evolved compared to nearby 
bright spirals. This paper presents
low frequency observations of 3 galaxies; PGC~045080 at 1.4 GHz, 610 MHz, 325MHz,
UGC~1922 at 610 MHz and UGC~6614 at 610 MHz. The observations were done with the GMRT.
Radio cores as well as extended structures were detected and mapped in all 
three galaxies; the 
extended emission may be assocated with jets/lobes associated with 
AGN activity.
Our results indicate that although these galaxies are optically dim, their nuclei
can host AGN that are bright in the radio domain. 
%% add here a maximum of 10 keywords, to be taken form the file <Keywords.txt>
\end{abstract}

\firstsection % if your document starts with a section,
              % remove some space above using this command.
\vspace{-3mm}
\section{AGN in Giant LSB Galaxies}

LSB galaxies have diffuse stellar disks, large HI gas disks and low metallicities
(Impey \& Bothun 1997). Although rich in gas they are poor in star formation and 
appear less evolved compared to bright galaxies. Their lack of evolution may be 
due to the presence of massive dark halos that inhibit the formation of disk 
instabilities such as bars and spiral arms, which can trigger
star formation activity in galaxies.
Active Galactic Nuclei (AGN) have been detected at optical wavelengths in 
several bulge-dominated giant LSB galaxies (Sprayberry et al. 1993; Schombert 1998).
This is suprising as AGN are generally associated with bright, star forming
galaxies (Ho, Philipenko \& Sargent 1997). Not much is known about the radio 
continuum emission from LSB galaxies.
Several giant LSB galaxies such as UGC~1922 are bright 
in the NVSS VLA survey at 1.4 GHz (Condon et al. 1998), and a
millimeter continuum source was detected in UGC~6614. These
observations suggest that AGN in giant LSBs have properties similar to those found in 
bright galaxies even though the galaxy evolutionary histories are very different.

\begin{figure}
 \includegraphics[height=1.7in,width=5.35in,angle=0]{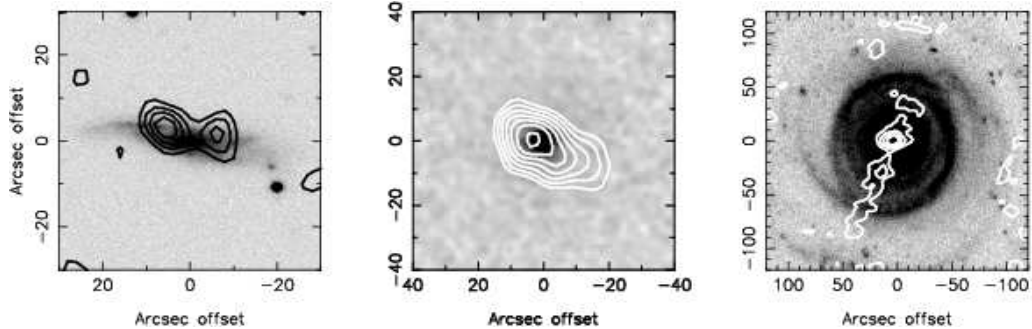}
  \caption{The panel shows 610~MHz observations of all three galaxies. From left to right
(a)~PGC~045080: Contours levels are 2, 3, 4, 5 times the noise level where the map noise is 
$\sigma=0.12~mJy~beam^{-1}$ and beam is ~7\sec; (b)~UGC~1922: Contours levels are 12, 14,
16, 18, 20, 22, and 24~$\sigma$ where $\sigma=0.49~mJy~beam^{-1}$ and the beam 
is ~ 24\sec; (c)~UGC~6614: Contours levels are 3, 5, 7, 9 and 11~$\sigma$   
where $\sigma=0.25~mJy~beam^{-1}$ and the beam is ~ 15\sec.}
\end{figure}

\vspace{-5mm}
\section{Galaxy Sample}

This poster presents preliminary results of a larger study of giant 
LSB galaxies using the GMRT. 
(i)~PGC~045080 is close to edge on and fairly distant
($v_{sys}=12,264$~\kms). Early optical studies did not detect an AGN in this galaxy (Sprayberry 
et al. 1993) but weak AGN activity may be present (Das et al. 2007). The galaxy is poor in star 
formation, fairly isolated and has a lopsided, massive HI disk. (ii)~UGC~1922 has a bright nucleus
or bulge and a very low surface brightness disk. It is also fairly distant ($v_{sys}=10,894$~\kms).
It is one of the rare LSB galaxies that have been detected in CO and there is a significant concentration
of molecular gas in the galaxy nucleus (O'Neil \& Schinnerer 2003). The galaxy hosts an AGN that is 
visible in optical emission as well as radio continuum. 
(iii)~UGC~6614 is a relatively nearby LSB galaxy ($v_{sys}=6,351$~\kms).
It is close to face on, has a prominent bulge and fairly distinct spiral arms that extend well 
into the disk. The AGN is visible in optical emission and appears as a compact core in the NVSS map
and at millimeter wavelengths as well (Das et al. 2006).

\vspace{-5mm} 
\section{GMRT Observations and Results}

We observed all three galaxies from August, 2005 to March, 2006 using the GMRT, which
is an array of thirty radio antennas arranged in a compact core and Y shaped configuration.
(i)~PGC~045080 was observed at 1.4~GHz, 610~MHz and 325~MHz (Das et al. 2007). The emission is
extended and at 610~MHz appears to have two lobes associated with
the nucleus (Figure~1a). The spectral index between 1.4~GHz and 325~MHz is $\alpha$=-0.63 
(where $S_{\nu}\propto\nu^{\alpha}$).
(ii)~The continuum emission at 610~MHz in UGC~1922 is also extended 
and the peak is offset from the galaxy center by a few arcseconds (Figure~1b). The spectral index 
between 1.4~GHz  and 610~MHz is 3 suggesting that it may be a Giga-Hertz Peaked Spectrum (GPS) radio
source. (iii)~The 610~MHz   
continuum map of UGC~6614 reveals a radio jet that extends well into the disk (Figure~1c). 
One side appears
brighter than the other and may represent the nearer jet. The core is bright and has a peak 
brightness of 3~mJy at 610~MHz. The spectral index is flat above 1.4~GHz but at 610~MHz is -0.53. 

\vspace{-5mm}
\section{Conclusions}

We have found extended radio continuum emission associated with AGN activity in 3 giant LSB 
galaxies. In at least two cases these represent radio jets/lobes. Thus though these galaxies
are optically dim, their nuclei can host AGN and associated energetic activities.  

\vspace{-1mm}
\begin{acknowledgments}

We thank the GMRT staff for help in the observations. 
The GMRT is operated by the National Center for Radio Astrophysics of 
the Tata Institute of Fundamental Research. We have used 
an SDSS image of PGC045080a and a 2MASS image of UGC1922. We thank 
Alice Quillen for the R band image of UGC6614. This work has made use 
of the NASA/IPAC Science Archive. 
\end{acknowledgments}

\end{document}